\documentclass[11pt]{article}
\usepackage{moriond,epsfig}

\def\be{\begin{equation}}
\def\ee{\end{equation}}
\def\bea{\begin{eqnarray}}
\def\eea{\end{eqnarray}}

\begin{document}
\vspace*{4cm}
\title{RECONSTRUCTING THE RIGHT-HANDED NEUTRINO MASSES IN LEFT-RIGHT SYMMETRIC THEORIES}

\author{ P. HOSTEINS \footnote{hosteins@cea.fr} }

\address{Service de Physique Th\'eorique, CEA-Saclay\\
91191 Gif-sur-Yvette, France}

\maketitle\abstracts{}

\section{Introduction~: reviewing the Seesaw Mechanism}

A very popular mechanism to generate masses for the left-handed neutrinos is the so-called seesaw mechanism, which consists of coupling the neutrinos to heavy particles and integrating them out.\\

The most studied type of seesaw is the type I~: we introduce 3 right handed neutrinos $N_R$ (singlets of $SU(2)_L$) with a heavy Majorana mass matrix $M_R$ and a usual Dirac coupling to the $\nu_L$ and the Higgs. When the Higgs gets a vev $v$ we generate a mass for the light neutrinos~:

\be
m_\nu=-v^2Y_\nu^T M_R^{-1}Y_\nu
\ee

There is also a type II seesaw, which introduces a scalar $\Delta_L$ triplet of $SU(2)_L$. We can write a coupling $\frac{1}{2}f_{Lij}l_{Li}^T\Delta_L Cl_{Lj}$ with $f_L$ symmetric. When the electrically neutral component of $\Delta_L$ gets a vev $\langle\Delta^0\rangle=v_L$ we get a mass for the neutrinos~:

\be
m_\nu=v_Lf_L
\ee

In some classes of models both types of contributions to the neutrino mass matrix are present. A frequent assumption is that one type of seesaw dominates over the other. We will consider that both types are present and solve for the heavy neutrino masses and mixings in this more general context. We then concentrate on the example of $SO(10)$.

\section{Reconstruction procedure of the heavy neutrino mass spectrum}

The general formula for the light neutrino masses we are considering is~:

\be
m_\nu=v_Lf_L-v^2Y_\nu^TM_R^{-1}Y_\nu
\ee

The theories which will allow us to solve for $M_R$ are theories which contain a left-right (LR) symmetry. In this context there exists a gauge group $SU(2)_R$ which is a mirror of $SU(2)_L$ for the right-handed particles and the heavy Majorana mass derives from the same mechanism as the type II seesaw, with a triplet $\Delta_R$ of $SU(2)_R$ . We have consequently $M_R=v_Rf_R$, with $\langle\Delta^0\rangle=v_R\gg v$, and the LR symmetry gives $f_L=f_R=f$ and $Y_\nu$ symmetric~:

\be
m_\nu=v_Lf-\frac{v^2}{v_R}Y_\nu^Tf^{-1}Y_\nu=\alpha f-\beta Y_\nu^Tf^{-1}Y_\nu
\ee

Now to extract f we need to know $Y_\nu$. Assuming $Y_\nu$ is known and we fix the neutrino low energy parameters, we can use $Z=Y_\nu^{-1/2}m_\nu(Y_\nu^{-1/2})^T$ and $X=Y_\nu^{-1/2}f(Y_\nu^{-1/2})^T$ to rewrite the above formula~:

\be
Z=\alpha X-\beta X^{-1} \qquad\qquad \mbox{Z and X symmetric}
\ee

\noindent where Z is known. We can then use a mathematical trick to find the matrix X (hence f)as a function of Z, namely we diagonalise both sides with a complex orthogonal transformation $O_Z=O_X$ \footnote{This transformation is not a physical diagonalisation and is not always well defined, so one should use it with caution}. The above equation then translates to each of the eigenvalues~:

\be
x_i^\pm=\frac{z_i\pm\sqrt{4\alpha\beta+z_i^2}}{2\alpha}
\ee

\noindent There are 2 solutions for each eigenvalue, which means a total of 8 solutions for f, in agreement with \cite{duality}. We label the solutions by the sign chosen for each $x_i$ for example ``+++''.

\section{The example of SO(10)}

We apply this procedure to supersymmetric $SO(10)$ theories with only symmetric mass matrices for the fermions and the following relations between the Yukawas~: $Y_\nu=Y_u$ and $Y_e=Y_d$ (which means that we have two ten-dimensional representations in the Higgs sector).

\subsection{Mass spectra}

\begin{figure}
\begin{center}
\psfig{figure=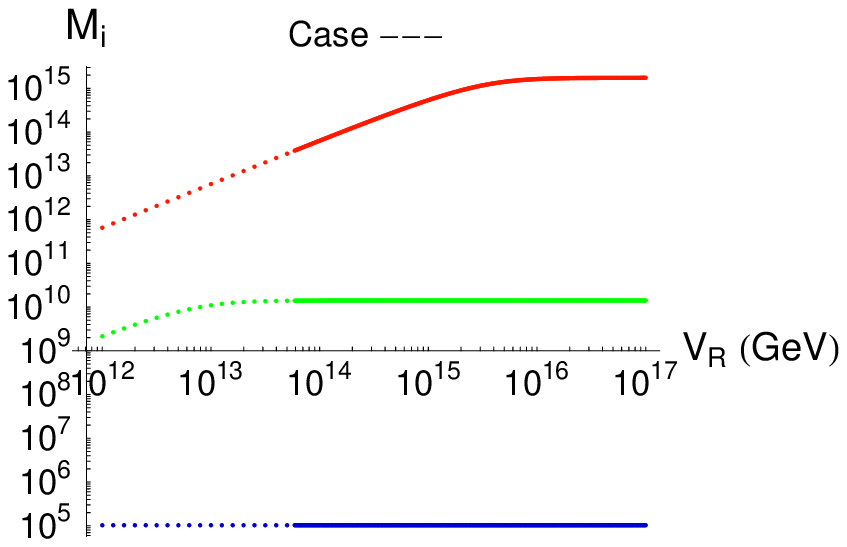,height=5.2cm}\hspace{1cm}
\psfig{figure=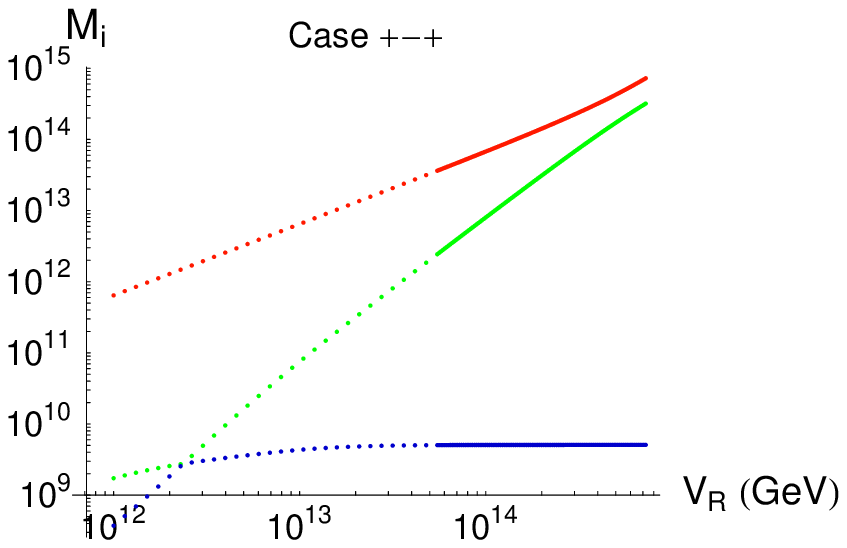,height=5.2cm}
\psfig{figure=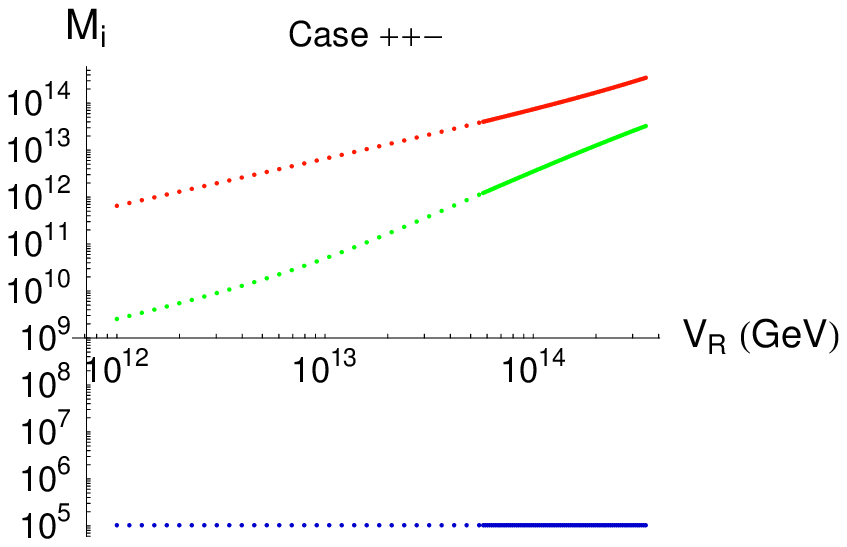,height=5.2cm}\hspace{1cm}
\psfig{figure=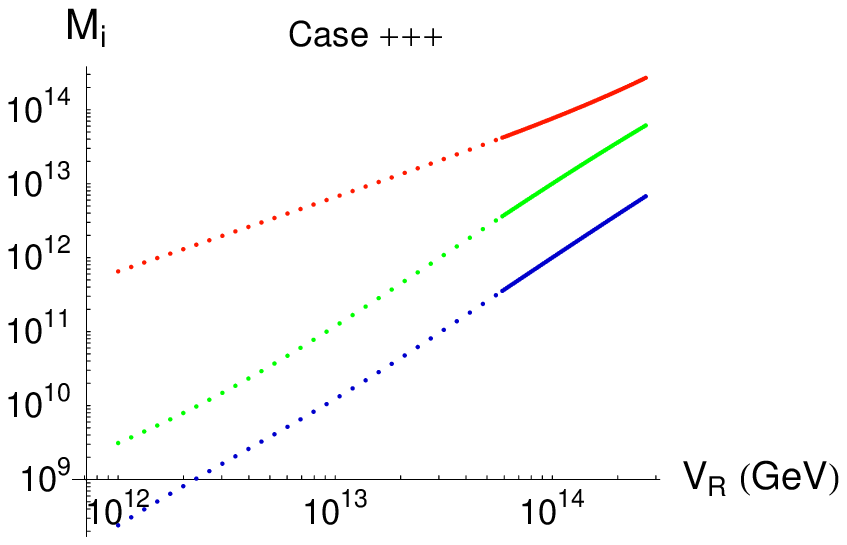,height=5.2cm}
\end{center}
\caption{Different mass spectra for the heavy neutrinos as a function of $v_R$ (no complex phase)~: the dotted part signals a tuning between the types I and II finer than 10\%
\label{masses}}
\end{figure}

As explained earlier we find 8 different mass spectra and a richer phenomenology than the $SO(10)$ type I seesaw alone. Examples of such mass spectra are displayed in fig.\ref{masses}.\\

\noindent Our free parameters are $v_R$ the B-L number breaking scale, at which the $N_{Ri}$ masses are generated, the ratio $\beta/\alpha$ and many complex phases as we cannot rephase the fermions independently.\\
We display the figures as functions of $v_R$ and fix $\beta/\alpha=1$. Although $v_R$ is allowed to run from $10^{12}$GeV to $10^{17}$GeV we cut the values for which the $f_i>1$ (perturbativity constraint). We also display dotted the region where the 2 contributions to $m_\nu$ begin to cancel each other with a fine-tuning smaller than 10\%.

\noindent 4 solutions have a very small $M_1\sim 10^5$GeV, just like in the pure type I $SO(10)$ seesaw, 2 have a quite stable $M_1\sim 10^{9-10}$GeV and the last 2 have a quite heavy spectrum.

\subsection{Leptogenesis}

Having the masses and couplings of the heavy neutrinos we can also explore the results for leptogenesis and try to see if we can accomodate the matter-antimatter asymmetry of the Universe.\\

\noindent First we review the mechanism of leptogenesis~: the $N_R$ having complex couplings they do not decay symmetrically to leptons and antileptons. Considering that heavy neutrinos are created in the hot early universe they start decaying at $T\sim M_i$ creating a lepton asymmetry. The asymmetry created by the lightest is driven by the quantity \cite{hambye,antusch}~:

\be
\varepsilon_{\mbox{\tiny CP}}\simeq\frac{3}{8\pi}\frac{\Im[Ym_\nu^*Y^T]_{11}}{(YY^\dagger)_{11}}\frac{M_1}{v^2}
\ee

\begin{figure}
\begin{center}
\psfig{figure=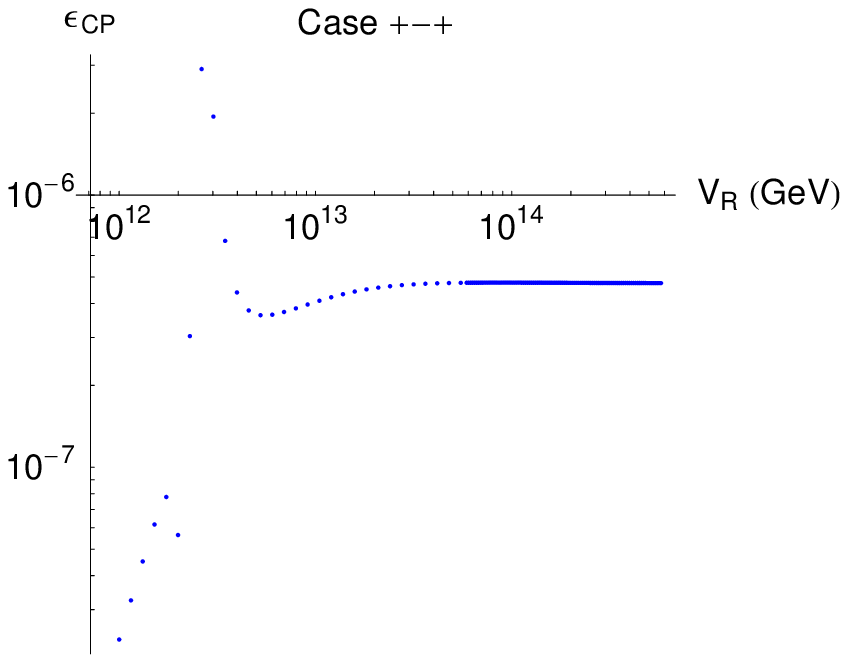,height=5.2cm}
\psfig{figure=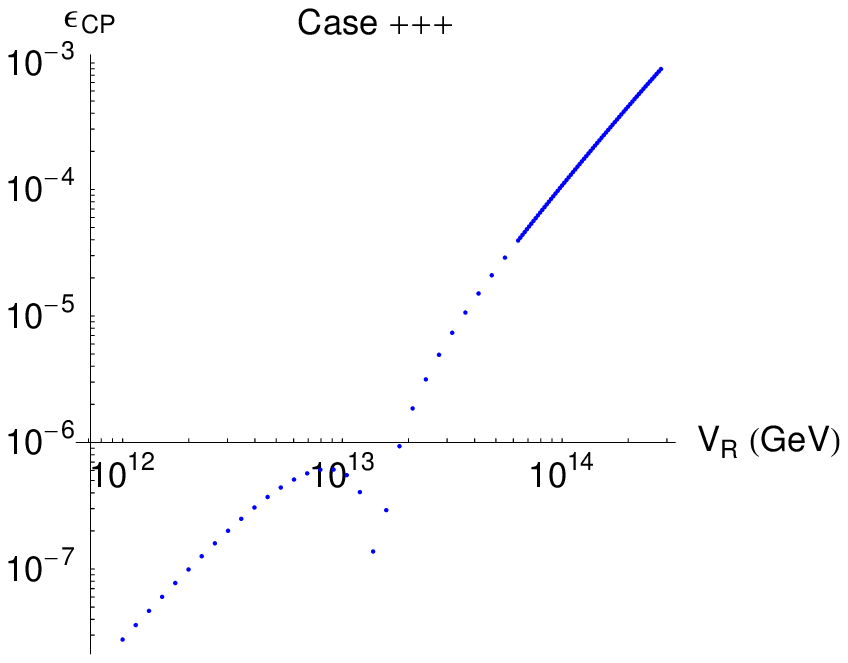,height=5.2cm}
\psfig{figure=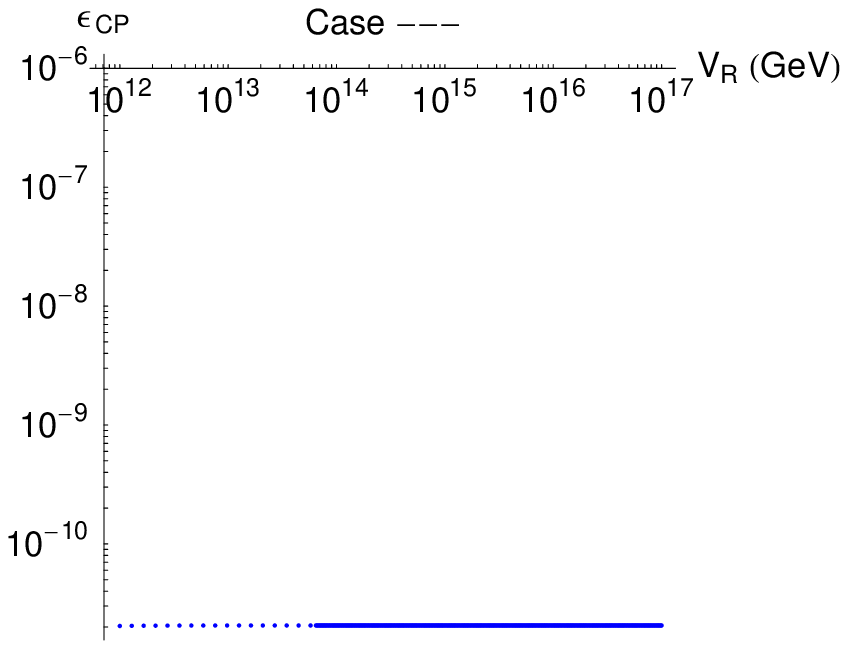,height=5.2cm}
\end{center}
\caption{The 3 typical behaviours of the CP asymmetry (one majorana complex phase is $\pi/4$, the others are 0)
\label{epsilon}}
\end{figure}

To find a good baryon asymmetry we must have $\varepsilon_{\mbox{\tiny CP}}> 10^{-6}$. We have computed $\varepsilon_{\mbox{\tiny CP}}$ for the different spectra and display generic examples in fig.\ref{epsilon}.

We see that some solutions are unable to reproduce a correct asymmetry, in connection with the smallness of $M_1$ as $\varepsilon_{\mbox{\tiny CP}}$ is proportionnal to it. Even taking into account flavour effects recently suggested \cite{vives,dibari}, that could be relevant for models with hierarchical $Y_\nu$, we are not able to accomodate sufficiently large values of $\varepsilon_{\mbox{\tiny CP}}$.\\
From the 4 remaining solutions, 2 of them can accomodate a good asymmetry for any value of the phases but they tend to favor heavy $M_1$ ($>10^{10}$GeV) which is in tension with constraints from gravitino production.\\
The last 2 give acceptable CP asymmetry for some specific values of the phases, including in the region with no tuning.

\section{Conclusion}

We have introduced a method to reconstruct the heavy neutrino mass matrix from the general typeI+typeII seesaw formula, postulating LR symmetry. In accordance with earlier work we found 8 different solutions, signing a richer phenomenology than pure type I or type II seesaw.\\
Applying the procedure to $SO(10)$, we found that, in contrast with type I $SO(10)$ it is possible to accomodate a correct baryon asymmetry.\\
Finally, it is worth mentionning that the large regions of fine tuning in $m_\nu$ can be naturally explained if we extend $SO(10)$ to the bigger gauge group $E_6$ where the couplings $Y_\nu$ and $f$ are exactly proportionnal at higher energies.

\section*{Acknowledgments}
I would like to thank St\'ephane Lavignac and Carlos Savoy for fruitful collaboration, as well as Michele Frigerio for useful advice. I am also grateful to the Moriond staff for allowing me to present this work and for the financial support.

\end{document}